\documentclass[12pt]{JHEP3}
\usepackage{amsmath,amssymb,bbm}
\usepackage{multirow}
\usepackage{youngtab}

\def\de{\partial}

\def\eq#1{(\ref{#1})}
\def\eqss#1#2{(\ref{#1}-\ref{#2})}
\def\eqs#1#2{(\ref{#1},\ref{#2})}
\def\2{\frac12}
\def\4{\frac14}
\def\ie{{\it i.e.}~}
\def\eg{{\it e.g.}~}

\newcommand{\be}{\begin{equation}}
\newcommand{\ee}{\end{equation}}
\newcommand{\bea}{\begin{eqnarray}}
\newcommand{\eea}{\end{eqnarray}}
\newcommand{\ba}{\begin{array}}
\newcommand{\ea}{\end{array}}
\def\a{\alpha}

\def\g{\gamma}

\def\d{\delta}
\def\e{\epsilon}

\def\l{\lambda}

\def\m{\mu}
\def\n{\nu}

\def\p{\pi}

\def\r{\rho}

\def\de{\partial}

\def\cH{{\cal H}}
\def\cM{{\cal M}}

\def\ad{{\dot \alpha}}

\author{Paul J. Heslop and Fabio Riccioni\\DAMTP,
Centre for Mathematical Sciences, University of Cambridge,
Wilberforce Road, Cambridge CB3 0WA,  UK\\ \email{P.J.Heslop,
F.Riccioni@damtp.cam.ac.uk}} 

\abstract{We discuss fermionic higher spin gauge symmetry breaking
in AdS space from a holographic perspective. Analogously to the
recently discussed bosonic case, the higher spin Goldstino mode
responsible for the symmetry breaking has a non-vanishing mass in
the limit in which the gauge symmetry is restored. This result is
precisely in agreement with the AdS/CFT correspondence, which
implies that ${\cal N}=4$ SYM at vanishing coupling constant
is dual to a theory in AdS which exhibits higher spin gauge symmetry
enhancement. When the SYM coupling is non-zero, the current
conservation condition becomes anomalous, and correspondingly the
local higher spin symmetry in the bulk gets spontaneously broken. We
also show that the mass of the Goldstino mode is exactly the one
predicted by the correspondence. Finally, we obtain the form of a
class of fermionic higher spin currents in the SYM side.}
\preprint{DAMTP-2005-62 \\ hep-th/0508086}


\title{On the fermionic {\it Grande  Bouffe}: more on higher spin symmetry breaking in
AdS/CFT}

\begin{document}

\section{Introduction}

The AdS/CFT correspondence~\cite{adscft} is a conjectured duality relating a
conformal field theory (CFT) in $d$ dimensions to a field theory
describing higher spin fields coupled to gravity in an AdS
background in $D=d+1$ dimensions. The most famous and remarkable example of this is the duality
between IIB superstring theory on $AdS_5 \times S^5$ with $N$ units of
5-form flux and $SU(N)$ ${\cal N}=4$ SYM theory in $d=4$. The
conjecture is most often 
discussed in the limit of large AdS radius where the higher spin
fields become extremely massive and decouple from the (super)gravity
modes.  In this limit the AdS side
is under control whereas the CFT side is not well understood since it
corresponds to 
the limit of large 't Hooft coupling. Therefore one can make
predictions for the strongly coupled CFT using
the correspondence but these can generally only be checked for certain
protected objects.

More recently the opposite limit, in which the CFT is weakly
coupled, has been discussed by a number of people~\cite{smallradius}
and it is possible to extrapolate the string spectrum and precisely
match it with the operator spectrum of free $\cal N$= 4 SYM in the
planar limit~\cite{bbms}. In
particular the limit of zero YM coupling has been conjectured to be
dual to a massless higher spin field theory which, although
inconsistent when coupled to gravity in flat space-time, can be
consistently defined in AdS spaces~\cite{Vasiliev90} (for a review
see~\cite{vasiliev} and references therein). When on the YM side the
coupling is turned on, in AdS all the massless higher spin fields
should develop a mass via the St\"uckelberg mechanism, essentially
by eating lower spin Goldstone fields. This phenomenon was termed
`La Grande Bouffe' in~\cite{bms}. The remaining massless fields
will all be contained in the supergravity multiplet. On the other hand,
in the dual CFT at zero coupling there will be infinitely many
higher spin conserved currents (in one-to-one correspondence with
the AdS higher spin gauge fields). The CFT counter-part of `La
Grande Bouffe' is the anomalous violation of these conserved
currents when the coupling is turned on, with the only remaining
conserved currents lying in the energy momentum multiplet. The most
famous example of this anomalous violation of classically conserved
currents is given by the Konishi multiplet.

The simplest example of a (bosonic) higher spin $s$ field is that of  a tensor
with $s$ completely symmetrised spacetime indices. For such an
object, in flat spacetime, the massless limit of a massive spin $s$ field
gives rise to $s+1$ massless fields of spins $0,1,\dots s$. However
the AdS/CFT correspondence predicts that the massless limit of a
massive spin $s$ field in AdS is a massless spin $s$ field  and a
massive spin $s-1$ field. The reason for this is that HS currents
$J_{i_1 \dots i_s}$ with $s
>2$ occur in ${\cal N}=4$ SYM, where they are conserved at vanishing
coupling $g=0$, and conformal invariance fixes the dimension of such
a spin $s$ conserved current on the $d$ dimensional boundary to be
$s+d-2$. Interactions are responsible
for their anomalous violation
$$ \de^{i_1} J_{i_1 \dots i_s} = g {\cal X}_{i_2 \dots
i_s}\label{anom} \quad ,
$$
and in the zero coupling limit the dimension of ${\cal X}$ is
$s+d-1$. This implies that ${\cal X}$ is not a conserved spin $s-1$
current when $g=0$, and therefore one expects it to be dual to a
massive field in the bulk.

In a recent paper~\cite{lagrandebouffe} the St\"uckelberg
formulation of bosonic massive higher spin fields (with completely
symmetrised spacetime indices) in AdS was derived.
The method consisted in obtaining the field equations in flat space
via dimensional reduction of the massless equations in one dimension
higher, and then including the necessary counterterms in order to
find the equivalent equations in AdS (see also~\cite{DESWALD,ZINO} for
similar
results using a different method). One can then use these equations
in AdS to extrapolate the massless limit, and one indeed obtains a
massless spin $s$ field and a massive spin $s-1$ field in line with
the CFT predictions (this phenomenon has also been discussed from a
cosmological viewpoint in~\cite{DESWALD} where it was termed `partial
masslessness'). The mass of the spin $s-1$ field one obtains in this
way is precisely the one predicted by AdS/CFT.

In this note we wish to continue this study by considering fermionic
higher spin currents.
We again try to covariantise the St\"uckelberg
formulation of massive fermionic higher spin fields from flat
space~\cite{tesifabio} to AdS space. We will only focus on higher spin
fermions whose spacetime indices are completely symmetrised.
The fermionic case turns out to be quite different from the bosonic
case, both in the auxiliary field structure
and in the fact that here the gauge transformations themselves
must be modified on covariantising from flat-space to AdS.

This
note is organised as follows. In section 2 we consider the
St\"uckelberg formulation of massive higher spin fermion field
equations in $D$-dimensional flat spacetime. The equations are
formally derived from $D+1$ dimensions, giving an exponential
dependence on the $D+1$-th coordinate to the field~\cite{tesifabio}.
In section 3 we consider the same equations in AdS, and we show that
gauge invariance requires the inclusion of additional counterterms
to the equations, such that one gets a massless spin $s$ field and a
{\it massive} spin $s-1$ field from a massive spin $s$ field in the
limit of zero mass. We also show that, exactly as in the bosonic
case, the mass of the spin $s-1$ field is precisely in agreement
with the one predicted by the AdS/CFT correspondence. Section 4
contains a discussion and the conclusions, focusing in particular on
the case in which $D=5$ (and $d=4$), corresponding to the duality
relating superstring theory on $AdS_5\times S^5$ to ${\cal N}=4$
SYM. All currents in the free theory are classified and examples of
the fermionic currents dual to the fermionic higher spin fields are
given together with their anomalous equations.

\section{Higher-spin fermions in flat spacetime}

In this section we study the St\"uckelberg formulation of massive
higher spin fermion field equations in flat spacetime. In analogy
with the bosonic case, analysed in \cite{lagrandebouffe}, we obtain
the massive equations in $D$ dimensions via dimensional reduction of
the massless equations in $D+1$ dimensions. We restrict our
attention to spinors that are completely symmetric with respect to
the spacetime indices. Reality properties of the spinors are not
relevant for this analysis. Moreover, in the interesting case of odd
$D$, we assume that the $D+1$ dimensional spinor is Weyl\footnote{In
  this case all $D+1$-dimensional $\gamma$-matrices in the equations
  below should be thought of as $D+1$ dimensional $\sigma$-matrices. }, so that
the $D$ dimensional one is Dirac. If instead $D$ is even, both the
$D+1$ and the $D$ dimensional spinors are Dirac. The massless
equations are
  \be
  \g\cdot\de\Psi_{M_1 \dots M_n} -n \ \de_{(M_1}(\g \cdot \Psi)_{M_2
\dots M_n)}
  =0\quad,
  \label{fermimassless}
  \ee
where the tensor-spinor $\Psi$ is completely symmetric with respect
to the spacetime indices, and satisfies the condition
  \be
  (\g\cdot\Psi)^M{}_M=0\quad. \label{3gamma}
  \ee
The equations are invariant with respect to the gauge transformation
  \be
  \d \Psi_{M_1 \dots M_n} = n \de_{(M_1} \e_{M_2 \dots M_n)}
  \quad , \label{gaugeD+1}
  \ee
where $\e$ is a spinor symmetric in its spacetime indices, and
satisfies the constraint
  \be
  \g\cdot\e = 0 \quad . \label{gammatrace}
  \ee
These equations can be obtained from the gauge-invariant lagrangian
\cite{FRONSDAL}
  \bea
  & &
  {\cal L} =\frac{1}{2}\bar{\Psi}^{(n)}_{...}\g^M \de_M \Psi^{(n)}_{...}+\frac{n}{2}
  \bar{\Psi}^{(n)}_{M ...}\g^M \g^N\de_N \g^P \Psi^{(n)}_{P ...}-\nonumber\\
  & & -\frac{n(n-1)}{8}\bar{\Psi}^{(n)L}{}_{L ...}
  \g^M \de_M\Psi^{(n)N}{}_{N...}+
  \nonumber\\
  \label{lagfer}
  & & + \frac{n(n-1)}{2}\bar{\Psi}^{(n)L}{}_{L...}
  \de_M \g_N \Psi^{(n)MN}{}_{...}- n \bar{\Psi}^{(n)}_{M...} \g^M
  \de_N \Psi^{(n)N}{}_{...} \quad .
  \eea

In order to obtain the equations for a massive field in $D$
dimensions in the St\"uckelberg formulation, we consider the field
$\Psi$ to depend harmonically on the $D+1$-th coordinate,
  \be
  \Psi_{\m_1 \dots \m_{n-k} y \dots y} (x,y) = (i)^k e^{- i \frac{\p}{4}
  \g_y} \psi^{(n-k)}_{\m_1 \dots \m_k}(x) e^{imy} + {\rm c. c.}
  \quad,
  \ee
where $\g_y$ is the gamma-matrix in the $D+1$-th direction. From eq.
(\ref{fermimassless}), we therefore obtain the equations
  \bea
  & & [ \g^\m \de_\m + m (1-k)] \psi^{(n-k)}_{\m_1 \dots \m_{n-k}} -
  (n-k) \de_{( \m_1 } (\g \cdot \psi^{(n-k)})_{\m_2 \dots \m_{n-k})}
  \nonumber \\
  & & -km (\g \cdot \psi^{(n-k+1 )})_{\m_1 \dots \m_{n-k}} - (n-k)
  \de_{(\m_1 } \psi^{(n-k-1)}_{\m_2 \dots \m_{n-k})} =0 \quad ,
  \label{n-keq}
  \eea
where $k=0,1,\dots , n$, and from the gauge transformations of eq.
(\ref{gaugeD+1}) one gets
  \be
  \d \psi^{n-k}_{\m_1 \dots \m_{n-k}} = (n-k) \de_{(\m_1}
  \e^{(n-k-1)}_{\m_2 \dots \m_{n-k})} + km \e^{(n-k)}_{\m_1 \dots
  \m_{n-k}} \quad , \label{gaugemassive}
  \ee
where the $D$-dimensional gauge parameters $\e^{(n-k)}$ are defined
by means of
  \be
  \e_{\m_1 \dots \m_{n-k-1} y \dots y } (x,y) = (i)^k
  e^{-i\frac{\p}{4} \g_y} \e^{(n-k-1)}_{\m_1 \dots \m_{n-k-1}} (x) e^{imy}
  + {\rm c.c.} \quad .
  \ee
In $D$ dimensions, the gamma-traceless condition of eq.
(\ref{gammatrace}) on $\e$ becomes
  \be
  \g^\m \e_\m^{(n-k)} + \e^{(n-k-1)} =0 \quad , \qquad k =
  1,\dots , n-1 \quad ,\label{epsilon}
  \ee
while the condition (\ref{3gamma}) on $\Psi$ becomes
  \be
  (\g \cdot \psi^{(n-k)})^\m{}_\m + \psi^\m{}_\m^{(n-k-1)} - (\g
  \cdot \psi )^{(n-k-2)} - \psi^{(n-k-3)} =0 \quad , \qquad k =
  0,\dots , n-3 \quad . \label{PsiD}
  \ee

If $m \neq 0$, one can use eq. (\ref{gaugemassive}) to gauge away
some of the lower rank spinors, while the lower rank spinors that
can not be gauged away because of the constraint (\ref{gammatrace})
turn out to be identically zero on shell, and are the auxiliary
fields of the massive theory. therefore, one ends up with an
equation for a massive spin $s=n+1/2$ field. In \cite{singhhagen} a
lagrangian formulation for spinors completely symmetric in their
spacetime indices was given in terms of a field $\psi^{(n)}$
satisfying the gamma-traceless condition $\g \cdot \psi =0$. The
condition that the divergence of this field vanishes on shell is
realised by means of a set of auxiliary fields $\psi^{(n-1)}$,
$\psi^{(n-k)}$, $\chi^{(n-k)}$, with $k =2 ,\dots , n$, all
satisfying a gamma-traceless condition. In \cite{tesifabio} it was
shown that these auxiliary fields are precisely the fields that can
not be gauged away in the St\"uckelberg formulation. Let us consider
for simplicity the case of spin $5/2$, {\ie} $n=2$. One can use the
gauge parameter $\e^{(1)}$ to gauge away $\psi^{(1)}$ completely,
while $\e^{(0)}$ can not be used because it is related to $\e^{(1)}$
via eq. (\ref{gammatrace}). We are therefore left with $\psi^{(2)}$
and $\psi^{(0)}$, and extracting the gamma-traceless part from
$\psi^{(2)}$ one gets exactly the auxiliary field structure of
\cite{singhhagen}. It is straightforward to see that a similar
analysis works for any $n$.

If $m=0$, the situation is different, since none of the gauge
parameters can be used to gauge away any of the fields, and
therefore all the fields $\psi^{(n-k)}$, $k=0,1,\dots , n$ become
massless.
In order to see this from our equations, one has to
perform recursive field redefinitions, so that eq. (\ref{epsilon})
becomes a gamma-traceless condition for the redefined parameters, while
eq. (\ref{PsiD}) becomes a condition of the form~\eq{3gamma}
for
the redefined fields, whose gauge transformations look exactly like
eqs. (\ref{gaugemassive}) with $m=0$ in terms of the new
parameters.

As we will see in the next section, and analogously to the
bosonic case, in AdS the massless limit of the St\"uckelberg
equations gives a completely different result, since imposing
masslessness for $\psi^{(n)}$ will result in a mass term for
$\psi^{(n-1)}$.

\section{Higher-spin fermions in AdS}

In this section we want to consider the AdS equivalent of the flat
space St\"uckelberg equations of the previous section for $m \neq
0$, and then study the limit of vanishing $m$. In analogy with the
bosonic case \cite{lagrandebouffe} and for consistency with
holography, we expect that the gauge fixing procedure that leads to
an equation for a massive spin $s$ field when $m \neq 0$ can still
take place up to spin $s-1$ when $m=0$, so that the resulting
equations in this limit describe a massless spin $s=n +1/2$
fermionic field $\psi^{(n)}$ and a {\it massive} spin $s-1 = n -
1/2$ fermionic field $\chi^{(n-1)}$.

In analogy with the bosonic case \cite{lagrandebouffe}, we consider
eq. (\ref{n-keq}) for $k=1$, and we gauge away $\psi^{(n-2)}$ using
$\e^{(n-2)}$. Therefore, only $\psi^{(n)}$ and $\psi^{(n-1)}$ (that
will be denoted with $\chi^{(n-1)}$ from now on) will appear in the
equation, and the only gauge invariance left is the one with respect
to the {\it gamma-traceless} parameter $\e^{(n-1)}$. We will see
that this gauge invariance in AdS will require the addition of a
mass term for $\chi^{(n-1)}$. We will then consider the limit of
vanishing $m$, and we will see that this mass term differs from the
AdS mass term\footnote{Recall that in AdS one defines a field to be
  massless if its equation of motion is gauge invariant. As we will
  see, such an
  equation contains a mass-like term which we call the AdS mass.}
for $\chi^{(n-1)}$. This result implies that the field
$\chi^{(n-1)}$ is massive in the limit in which $\psi^{(n)}$ becomes
a massless field.

We now derive first the AdS mass for a generic spin $s= l +1/2$
field. We therefore consider the equation
  \be
  \g\cdot\nabla \psi_{\m_1 \dots \m_l} -l \ \nabla_{(\m_1}(\g \cdot \psi)_{\m_2
  \dots \m_l)} + M_{AdS} \psi_{\m_1 \dots \m_l} + \tilde{M}_{AdS}
  \g_{(\m_1 } (\g \cdot \psi )_{\m_2 \dots \m_l )}
  =0\quad,
  \label{fermimasslessads}
  \ee
where $\psi^{(l)}$ satisfies the constraint $(\g \cdot
\psi^{(l)})^\m{}_\m =0$, and its gauge transformation is
  \be
  \d \psi^{(l)}_{\m_1 \dots \m_l } = l \nabla_{(\m_1}
  \e^{(l-1)}_{\m_2 \dots \m_l )} + M' \g_{(\m_1 } \e^{(l-1)}_{\m_2 \dots
  \m_l )} \quad , \label{gaugeads}
  \ee
  where $\e$ is gamma-traceless,
and we want to determine $M_{AdS}$, $\tilde{M}_{AdS}$ and $M'$ such
that eq. (\ref{fermimasslessads}) is gauge invariant with respect to
the transformation of eq. (\ref{gaugeads}). Using the fact that the
commutator of two covariant derivatives acting on a spinor with
$l-1$ vector indices is
  \be
  [ \nabla_\m , \nabla_\n ] \e_{\r_1 \dots \r_{l-1}} =
  - \frac{1}{2 L^2} \g_{\m\n} \e_{\m_1 \dots \m_{l-1}} +
  \frac{l-1}{L^2} ( g_{\n (\r_{1}} \e_{\r_2 \dots \r_{l-1}) \m} -
  g_{\m ( \r_1} \e_{\r_2 \dots \r_{l-1}) \n} ) \quad ,
  \ee
where $L$ is the AdS radius, one obtains
  \be
  M' = \frac{l}{2L} \quad ,
  \ee
  \be
  M_{AdS} = \frac{1}{2L} [ D + 2 (l-2)] \quad , \label{relevantmass}
  \ee
  \be
  \tilde{M}_{AdS} = \frac{l}{2L} \quad .
  \ee
Since the gamma-trace of the field can always be put to zero
choosing a suitable gauge, the relevant AdS mass term is $M_{AdS}$
in eq. (\ref{relevantmass}).

We now consider the equation for the spin $n+1/2$ field $\psi^{(n)}$
and the spin $n-1/2$ St\"uckelberg field $\chi^{(n-1)}$, after
having gauged away $\psi^{(n-2)}$ by means of $\e^{(n-2)}$. The
covariantisation of eq. \eq{n-keq} for k=0 takes the form of eq.
\eq{fermimasslessads} together with mass terms for $\psi^{(n)}$ and
a term involving the derivative of the spin $n-1/2$ St\"uckelberg
field $\chi^{(n-1)}$,
\begin{eqnarray}
  &&\g\cdot\nabla \psi^{(n)}_{\m_1 \dots \m_n} -n \ \nabla_{(\m_1}(\g \cdot
  \psi^{(n)})_{\m_2
  \dots \m_n)} + M_{AdS} \psi^{(n)}_{\m_1 \dots \m_n} + \tilde{M}_{AdS}
  \g_{(\m_1 } (\g \cdot \psi^{(n)} )_{\m_2 \dots \m_n )}\nonumber \\
  &&+m\,\psi^{(n)}_{\m_1 \dots \m_n}-n \nabla_{(\m_1 } \chi^{(n-1)}_{\m_2 \dots
  \m_n
  )} - \frac{n}{2L} \g_{(\m_1 } \chi^{(n-1)}_{\m_2 \dots \m_n )}
  =0\quad.\label{psineq}
\end{eqnarray}
This equation is gauge invariant under
  \be
  \d \psi^{(n)}_{\m_1 \dots \m_n} = n \nabla_{(\m_1 } \e_{\m_2 \m_n
  )} + \frac{n}{2L} \g_{(\m_1 } \e_{\m_2 \dots \m_n )} \quad ,
  \qquad \d \chi^{(n-1)}_{\m_1 \dots \m_{n-1}} = m \e_{\m_1 \dots
  \m_{n-1}} \quad .\label{gtrans}
  \ee
Indeed the first and second lines are separately gauge invariant:
the first line is simply the equation for a massless spin $n+1/2$
field in AdS derived above, whereas in the second line there is a
mass term for $\psi^{(n)}$ whose gauge variation is cancelled by the
gauge variation of the terms involving $\chi^{(n-1)}$.

In order to derive the equation for $\chi^{(n)}$, we consider the
expression that we get from eq. (\ref{n-keq}) for $k=1$,
  \be
  \g^\m \nabla_\m \chi^{(n-1)}_{\m_1 \dots \m_{n-1}} - (n-1)
  \nabla_{(\m_1} (\g \cdot \chi^{(n-1)})_{\m_2 \dots \m_{n-1})} - m
  (\g \cdot \psi^{(n)})_{\m_1 \dots \m_{n-1}}\quad .
  \ee
Its variation with respect to the gauge transformation~\eq{gtrans}
with $\e$ gamma-traceless, is
  \be
  - \frac{m}{2L} [ D+ 2 (n-1)] \e_{\m_1 \dots \e_{n-1}} \quad ,
  \ee
and in order to cancel this, one needs to include the mass term
  \be
  M \chi^{(n-1)} = \frac{1}{2L} [ D+2 (n-1)] \chi^{(n-1)} \quad ,
  \ee
so that the resulting equation is
  \bea
  & & \g^\m \nabla_\m \chi^{(n-1)}_{\m_1 \dots \m_{n-1}} - (n-1)
  \nabla_{(\m_1} (\g \cdot \chi^{(n-1)})_{\m_2 \dots \m_{n-1})} \nonumber \\
  & & - m (\g \cdot \psi^{(n)})_{\m_1 \dots \m_{n-1}} +
  \frac{1}{2L} [ D+2 (n-1)] \chi^{(n-1)}_{\m_1 \dots \m_{n-1}} =0
  \quad .\label{chieq}
  \eea
If $m \neq 0$, one can still gauge away the St\"uckelberg field
$\chi^{(n-1)}$, so that only a massive spin $s$ field remains. If
$m=0$, though, it is clear from eq. (\ref{gtrans}) that this is no
longer possible. What we are going to show now is that one can
consistently consider the $m \rightarrow 0$ limit of
Equations~\eqs{psineq}{chieq}, since all the lower rank fields can
still be gauged away in this limit.

Indeed, the fact that $M \neq M_{AdS}$ proves that in the limit of
vanishing $m$, in which the field $\psi^{(n)}$ becomes massless, the
field $\chi^{(n-1)}$ remains a massive field. The difference between
the square of the mass $M$ and the square of the AdS mass $M_{AdS}$
for $\chi^{(n-1)}$, obtained from~\eq{relevantmass} with $l=n-1$, is
  \be
  M^2 L^2 - M_{AdS}^2 L^2 = 2 D +4 n - 8 \quad .
  \ee
This result is in perfect agreement with the AdS/CFT
relation~\cite{ferrara}
  \be
  M^2 L^2 - M_{AdS}^2 L^2 = \Delta ( \Delta - d ) - \Delta_{min} (
  \Delta_{min} -d ) \quad ,
  \ee
where $d = D-1$ and  $\Delta$ is the dimension of the operator dual
to $\chi^{(n-1)}$, whose value is
  \be
  \Delta = d + n - \frac{1}{2}
  \ee
in the limit of vanishing Yang-Mills coupling, while $\Delta_{min} =
d + n - \frac{5}{2}$ represents the conformal unitary bound for the
dimension of a spin $n- 1/2$ operator. Finally, as in the flat
spacetime case, a field redefinition for $\psi^{(n)}$ is needed in
order to recover the standard massless equation for a spin $s = n+
1/2$ field in AdS when $m=0$.

\section{Discussion and conclusions}

Free ${\cal N}=4$ SYM contains many spin $s$ conserved currents
which conformal representation theory tells us must have dimension
$\Delta=s+2$. The fundamental fields of the theory,
$\phi_I,\l_{\a}^{A},\bar \l_{\ad A},F_{ij}$ all have twist
($\Delta-s$) one, and since any composite operator must have twist greater
than or equal to the sum of the twists of its constituent fundamental
fields in the free theory, it follows that all currents must be bilinear in the
fundamental fields and are thus elements of the higher spin
doubleton multiplet~\cite{sezsund}. Indeed it turns out that
all bilinear conformal primary operators have twist two and so the
following statement is true:
all currents belong to the higher spin doubleton multiplet and
conversely all
primary operators in the doubleton multiplet with spin $j_1>0,\ j_2>0$
are currents.

The higher spin doubleton multiplet splits into infinitely many
representations of the superconformal group, namely the
energy-momentum supermultiplet ${\cal T}_{IJ}:= Tr(W_I W_J - 1/6
\d_{IJ}W_KW_K)$, the Konishi supermultiplet ${\cal H}:= Tr(W_I W_I)$
and the higher spin analogues of the Konishi
supermultiplet~\cite{KONISHI}
$\cH^{(s)}\sim Tr(W_I \de^s W_I) +\dots, \quad s\in 2 {\mathsf{ Z
\hspace{-1ex} Z }}$ (the precise form of the latter are given
in~\cite{lagrandebouffe} where one can also find details about our
notation). Since all component currents in $N=4$ SYM
are contained in these multiplets they are all of the
form
\begin{equation}
  D_{(\a_1 A_1} \dots D_{\a_k A_k} \bar D_{(\ad_1}^{A_1} \dots \bar
  D_{\ad_l}^{A_l} \ \cH^{(s)}_{\a_{k+1}\dots
  \a_{k+s})\ad_{l+1}\dots\ad_{l+s}}|_{\theta=\bar \theta=0} \quad
  0\leq k,l \leq 4 \label{compcurrents}
\end{equation}
or are currents contained in the energy-momentum multiplet.
One can
thus straightforwardly classify all currents in the theory (see the
table below and also~\cite{osborn}).

\begin{table}[h]
$$\begin{array}{|c|c|c|c|c|c|} \hline
\mbox{Current} & \mbox{Spin} (j_1,j_2)\rule[-1.9ex]{0pt}{5ex}&
\cH^{(s)}:s= & (k,l)& SU(4) \mbox{rep.}\\ \hline
 \multirow{5}{*}{$J_{(i_1\dots i_{n})}$} & \multirow{5}{*}{$({n\over 2},{n\over 2})$}
&   n      & (0,0)&{\bf 1}                \\
   &   &n-1      & (1,1)&
{\bf
4\times\bar 4 }\rule[-1.3ex]{0pt}{3.8ex}\\
&            &   n-2    & (2,2)& {\bf 6\times6 }\\
&     &   n-3      & (3,3)& {\bf \bar 4\times 4 }\\
&                &   n-4    & (4,4)& {\bf 1   }\\  \hline
\multirow{4}{*}{ $\Psi_{\ad (i_1\dots i_{n})}$} &  \multirow{4}{*}{
$(\frac n2,\frac{n+1}2) $}       &   n      & (0,1)& {\bf \bar 4
}\rule[-1.3ex]{0pt}{3.8ex}\\
  &
   &   n-1      & (1,2)& {\bf
4\times 6 }\\
&     &   n-2      & (2,3)& {\bf 6\times  4 }\\

 &     &   n-3      & (3,4)& {\bf \bar 4 }\\\hline
 \multirow{3}{*}{$J_{[kj](i_1\dots i_{n})}$} & \multirow{3}{*}{ $ (\frac n2,\frac n2+1)$}
&   n      & (0,2)& {\bf 6 }\\
 &
\rule[-1.3ex]{0pt}{3.8ex}     &   n-1      & (1,3)& {\bf 4\times 4
}\\
&     &   n-2      & (2,4)& {\bf 6 }\\
\hline
  \multirow{2}{*}{$\Psi_{\ad [jk](i_1\dots i_{n})}$} &  \multirow{2}{*}{$(\frac n2,\frac{n+3}2)$}   \rule[-1.3ex]{0pt}{3.8ex}
&   n      & (0,3)& {\bf  4 }\\
 &
\rule[-1.3ex]{0pt}{3.8ex}      &   n-1      & (1,4)& {\bf  4
}\\\hline
 J_{[jk][lm](i_1\dots i_{n})} &  (\frac n2,\frac n2+2)     \rule[-1.3ex]{0pt}{3.8ex}   &
n      & (0,4)& {\bf 1 }\\\hline
\end{array}
$$
\caption{}
{\small Table giving all currents (up to complex
conjugation) in $N=4$ SYM (apart from
  those in the energy-momentum multiplet), the supermultiplets
  they appear in and the $SU(4)$ representations they carry. The
  numbers $(k,l)$
  indicate where the currents lie in the supermultiplet according
  to~\eq{compcurrents}. These operators are only currents for $j_1>0, \ j_2>0$.
  We must also have $s\geq0$ and even. The mixed symmetry tensors lie in irreducible
  representations of the Lorentz group given by Young
  tableaux~\eqss{yt1}{yt3}. Finally, the complete current content of $N=4$
  SYM must include the
  energy-momentum multiplet which contains 1 real spin $(1,1)$ current,
  $\bf 4$ complex spin $(1,1/2)$ currents, and $\bf 15$ spin $(1/2,1/2)$
currents.}

\end{table}

In the interacting $N=4$ SYM theory, all higher spin currents, other
than those in the energy-momentum supermultiplet, become anomalous
meaning their current conservation condition is violated $\de^{i_1}
J_{i_1 \dots i_s} = g {\cal X}_{i_2 \dots i_s}$.  The anomalous
(non)-conservation equations for these currents are encoded in
analogous anomalous equations for the $N=4$ superfields
$\cH_{\a_1\dots\a_s\ad_1\dots\ad_s}$ (supercurrents) containing
these currents. The relevant superfield equations are
\begin{eqnarray}
D^{\a}_A{\cal H}^{(s)}_{\a...}=g {\cal M}_{A ...} &\quad& \bar
D^{\ad A} {\cal H}^{(s)}_{\ad ...}=g \bar{\cal M}^{A}_{...}\qquad
s\geq 2\label{sanom1}\\
D_{\a A} D_B^{\a}{\cal H}^{(0)}=g {\cal M}_{AB} &\quad& \bar
D_{\ad}^{A} D^{\ad B}{\cal H}^{(0)}=g \bar{\cal M}^{AB}\ .
\end{eqnarray}
Here $\cal H,\, M, \bar M$ are separate $N=4$ supermultiplets in
the free theory ($g=0$) whereas in the interacting theory we can see
that they combine and indeed they further combine with another
supermultiplet $\cal N$ to form one long supermultiplet\footnote{The precise
  forms of $\cal M, \bar M$ and $\cal N$ are given in
  \cite{lagrandebouffe,Henn:2005mw}.}. 

In~\cite{lagrandebouffe} the St\"uckelberg AdS equations for massive
bosonic spin $s$ fields, completely symmetric in their spacetime
indices, was given. These fields are the holographic duals of the
currents $J_{(i_1 \dots i_n )}$ in the first row of table 1. In the
limit of zero mass for the spin $s$ field, a massive spin $s-1$
field appears, whose mass is exactly in agreement with the AdS/CFT
correspondence. In this note we have studied the same mechanism for
the fields dual to the fermionic currents $\Psi_{\ad (i_1\dots
i_{n})}$ in the second row of table 1. The simplest example of such
fermionic currents is the first supersymmetric descendant of the
higher spin analogues of the Konishi supermultiplet,
\begin{equation}
\l^{(s)}_{A\a_0\dots\a_s\ad_1\dots\ad_ s}:=D_{(\a_0 A} \cH^{(s)}_{\a_1
  \dots \a_s)\ad_1\dots \ad_s}\ .
\end{equation}
This is the descendant with $(k,l)=(1,0)$ (the complex conjugate of
the spin $(n/2,n/2+1/2)$ current in  the $\bf \bar 4$ representation in table 1) and the violation of
the current conservation equation can be found using standard
superfield methods from the above superfield equations~\eq{sanom1} to be
\begin{eqnarray}
4\de^{\a_0\ad_0}\l_{\a_0\dots\a_s\ad_0\dots\ad_{s-1}}&=&
g/2\de_{(\a_1}{}^{\ad_0} \cM_{B\a_2\dots\a_s)\ad_0\dots\ad_{s-1}}|_{\theta=\bar\theta=0}\nonumber\\
&&-igD_{(\a_0 B}D_{A}^{\a_0}\bar\cM^A_{\a_1\dots\a_{s}\ad_1
\dots\ad_{s-1}}|_{\theta=\bar\theta=0}\\
&&+ig/2[\bar D^{\ad_0 A},D_{(\a_1 B}]
\cM_{A\a_2\dots\a_s)\ad_0\dots\ad_{s-1}}|_{\theta=\bar\theta=0}\nonumber \ .
\end{eqnarray}
As in the bosonic case, we have shown that a massive fermion field
of spin $s$ in AdS becomes a massless spin $s$ field and a massive
spin $s-1$ field. We computed the mass of the spin $s-1$ field,
showing that it exactly agrees with the dimension of the dual
operator in SYM in the limit of vanishing coupling.

It would be interesting to study the same mechanism for the fields
dual to the remaining currents in table 1. The Young tableaux
associated to them are
  \be
  \overbrace{\yng(5,1)}^{n+1} \quad ,\label{yt1}
  \ee
  \be
  \overbrace{\yng(5,1)}^{n+1}{}^\a \quad ,
  \ee
  \be
  \overbrace{\yng(5,2)}^{n+2}\quad ,\label{yt3}
  \ee
corresponding respectively to the last three rows of the table. The
field strengths of these fields satisfy self-duality conditions that
are generalisations of the self-duality condition of the
antisymmetric 2-forms in the supergravity multiplet~\cite{sezsund}. Massive fields of mixed symmetry in $AdS_5$ were discussed
in \cite{metsaevmixsym}, while flat space equations for gauge fields
in arbitrary representations of the Lorentz group were introduced in
\cite{siegzwie} and more recently discussed in \cite{bekboul}.
We hope to soon report on progress in this direction.

\vskip 2cm

\section*{Acknowledgments}
We are grateful to I. Buchbinder, F. Dolan and in particular to M. Bianchi for
discussions. The work of F.R. is supported by a European Commission
Marie Curie Postdoctoral Fellowship, Contract MEIF-CT-2003-500308.

\end{document}